\documentstyle[preprint,aps,amsmath]{revtex}
\begin{document}

\draft

\title{ A Scattering method to the equilibrium spin current in a ferromagnet junction}

\author{J. Wang$^{1,2}$ and K.S. Chan$^{1}$}
\address{$^1$Department of Physics and Materials Science, City University of Hong Kong, Tat Chee
Avenue, Kowloon, Hong Kong, P.R. China}
\address{$^2$Department of Physics, Southeast University, Nanjing
210096, P.R. China}
\date{\today}
\maketitle

\begin{abstract}
We extended McMillan's Green's function method to study the
equilibrium spin current (ESC) in a ferromagnet/ferromagnet
(FM/FM) tunnelling junction, in which the magnetic moments in both
FM electrodes are not collinear. The single-electron Green's
function of the junction system is directly constructed from the
elements of the scattering matrix which can be obtained by
matching wavefunctions at boundaries. The ESC is found to be
determined only by the Andreev-type reflection amplitudes as in
the Josephson effect. The obtained expression of ESC is an exact
result and at the strong barrier limit gives the same explanation
for the origin of ESC as the linear response theory, that is, ESC
comes from the exchange coupling between the magnetic moments of
the two FM electrodes,
${\mathbf{J}}\sim{\mathbf{h}}_{l}\times{\mathbf{h}}_{r}$. In the
weak barrier region, ESC cannot form spontaneously in a
noncollinear FM/FM junction when there is no tunneling barrier
between the two FM electrodes.
\end{abstract}

\section{introduction}
Spin related transport in magnetic hybrid systems has been studied
intensively for last two decades and considerable progress has
been achieved since the discovery of the tunneling
magnetoresistance (TMR) effect\cite{1} in ferromagnet (FM)
tunnelling junctions. This effect originates from the electrical
resistances of these junction being dependent strongly on whether
the moments of adjacent magnetic layers are parallel or
antiparallel. The reason is that the tunnelling electrons are
scattered more strongly in an antiparallel magnetic configuration
than in a parallel configuration. As a result, tunnelling
junctions with moments in adjacent magnetic layers aligned
antiparallelly have larger overall resistance than when the
adjacent moments are aligned parallel, giving rise to TMR. The
reverse effect of TMR is the spin transfer effect predicted
independently by Slonczewski\cite{2} and Berger\cite{3} about a
decade ago, in which a sufficiently large spin-polarized current
injected from a normal metal (NM) into a FM layer can lead to
magnetic moment reversal in the FM layer.

Spin transfer torque occurs in magnetic multilayers with
noncollinear moments and is due to the non-conservation of spin
current through the interface between NM and FM. Due to the
presence of noncollinearity, the component of spin current
transverse to the magnetization of the layers is not transmitted
across the interfaces between NM and FM; in other words, the
discontinuity of spin current at the interface is the origin of
the spin transfer effect that can result in magnetization
precession or reversal.\cite{4,5,6,7,8} A large number of
experiments have observed this spin transfer effect in magnetic
multilayer structures,\cite{9,10,11} in which one of the FM layers
is very thick and works as a polarizer of electric current with
its fixed moment, and the moment of the thin FM layer may switch
when a strong polarized electric current perpendicular to the
layer plane flows through the layers.

As the two magnetization directions in the FM/FM tunnelling
junction are misaligned, an equilibrium spin current (ESC) can
flow in the junction without any bias,\cite{7} this phenomena is
analogous to the Josephson effect that the superconductor
macroscopic phase difference between the two sides of a junction
drives a supercurrent through the junction. The dissipationless
ESC will disappear if two moments in the adjacent FM layers are
collinear, parallel or antiparallel. The existence of ESC
$\mathbf{J}$ has been verified by many authors using the linear
response theory,\cite{12,13,14,15,16} and explained as the result
of the exchange coupling between two magnetic moments
${\mathbf{h}}_r$ and ${\mathbf{h}}_{l}$,
${\mathbf{J}}\sim{\mathbf{h}}_{l}\times{\mathbf{h}}_{r}$, Thus the
magnetization phase difference between the FM electrodes induces
ESC in the FM junction as in the Josephson effect.  In the case of
a thin barrier between FMs or the strong coupling limit, the
behavior of ESC is still unknown in the literature and worth
studying.

In this paper, we study the ESC flowing in a FM/FM junction with
two fixed non-collinear magnetic moment using a simple quantum
mechanical approach. The present work is motivated by two reasons.
Firstly, dissipationless spin current in last several years has
drawn considerable interest and also incurred much debate, such as
the controversies in the spin Hall effect.\cite{17,18,19,20} The
study of ESC in FM/FM junctions may shed light on the spin current
in spin-orbit coupled systems. Secondly, the well-known result of
ESC ${\mathbf{J}}\sim{\mathbf{h}}_{l}\times{\mathbf{h}}_{r}$ in
FM/FM junctions in the literature was obtained by using the linear
response theory\cite{12,13,14,15,16} or heuristic
derivation,\cite{21} thus an exact result of ESC is in order. By
extending McMillan's Green's function method which was originally
employed to study the Josephson effect in superconductor
junction,\cite{22} we obtained an expression of ESC in the FM/FM
junction at equilibrium. ESC is determined only by the
Andreev-type reflection amplitudes,\cite{8,23} which is similar to
the Josephson effect.\cite{24,25} It was also found that the exact ESC expression
is slightly different from the linear response result; in other
words, the high-order tunnelling processes are negligible and the
linear response approximation has a wider range of application.
When there is no barrier between the two FM electrodes, the ESC
was found to vanish indicating that ESC is a pure quantum
mechanical effect.

This paper is organized in the following way. In Section II, a
FM/FM junction model is given and the single electron Green's
function of the junction system is constructed from the scattering
coefficients. An analytic expression of ESC in the FM/FM junction
is obtained in the third section and some discussions are
presented. A conclusion is drawn in the last section.

\section{Green's function}
The FM/FM tunnelling junction is depicted schematically in Fig.~1 and
the layer between the FM electrodes can be either an insulator barrier
or a normal metal. A simple free electron model is adopted to
describe the FM/FM tunnelling junction and the Hamiltonian  reads
\begin{equation}
 {\cal H}=\frac{-{\hbar}^{2}{\nabla}^{2}}{2m_{e}}+V(x)-
 \theta(-x){\mathbf{h}}_{l}{\cdot}{\boldsymbol{\sigma}}-\theta(x-L){\mathbf{h}}_{r}
 {\cdot}{\boldsymbol{\sigma}},
 \end{equation}
where $m_e$ is the effective electron mass and assumed to be identical
in all three regions, the two FM electrodes and the middle region.
${\mathbf{h}}_{l}$ and ${\mathbf{h}}_{r}$ are the internal
molecular fields of the left and right FM, respectively.
${\boldsymbol{\sigma}}$ denotes the Pauli spin operator, and
$\theta(x)$ is the step function. The potential energy $V(x)$ may
take different values in different regions but remains constant in
both FMs. Here the molecular fields ${\mathbf{h}}_{l}$  and
${\mathbf{h}}_{r}$ (magnetization with energy unit) are not
collinear in our consideration and assumed to be fixed by an external
magnetic field or other methods. Without loss of generality, we
take the spin quantization axis of the system to be parallel to the
magnetization of the left FM ${\mathbf{h}}_{l}$ and the direction
of ${\mathbf{h}}_{r}$  of the right FM is described by the polar
coordinate ($\theta$, $\phi$).

In a free electron model, the energy dispersions of the two FMs can
readily be solved and they are spin-dependent. In the left FM, the
eigenvalue is $E_{\pm}=\frac{\hbar^{2}k^2}{2m_{e}}\pm h_{l}+U_{l}$
and the spinor is the eigenfunction of $\sigma_{z}$, and in the right
FM, $E_{\pm}=\frac{\hbar^{2}k^2}{2m_{e}}\pm h_{r}+U_{r}$ and the
spin eigenfunctions are $\psi_{+}^{l}=\left(%
\begin{array}{c}
  \cos(\theta/2)e^{-i\phi/2}  \\
  \sin(\theta/2)e^{i\phi/2}  \\
\end{array}%
\right)$ and $\psi_{-}^{l}=\left(%
\begin{array}{c}
  -\sin(\theta/2)e^{-i\phi/2}  \\
  \cos(\theta/2)e^{i\phi/2}  \\
\end{array}%
\right)$ where $U_{l} (U_{r})$ are the different potential energy
in the left(right) FM, $\mathbf{k}$ is the wavevector of electron and
$\pm$ is the spin index. The spatial eigenfunctions are plane
waves. Due to spin splitting, there are four incoming
wave functions with their corresponding outgoing wave functions
in the left and right FM, as schematically shown in Fig.~2.
$\Psi_{1}(x)$ and $\Psi_{2}(x)$ are the wavefunctions of the
minority spin and majority spin electron injecting from the left
FM, respetively, while $\Psi_{3}(x)$ and $\Psi_{4}(x)$ are those
injecting from the right FM. For example, the wavefunction of the
first type of scattering event $\Psi_{1}(x)$ is given by
\begin{equation}
\Psi_{1}(x)=
\left\{%
\begin{array}{l}
     \exp(ik_{+}^{l}x)\left(%
\begin{array}{c}
  1 \\
  0 \\
\end{array}%
\right)+a_{1}\exp(-ik_{+}^{l}x)\left(%
\begin{array}{c}
  1 \\
  0 \\
\end{array}%
\right)+b_{1}\exp(-ik_{-}^{l}x)\left(%
\begin{array}{c}
  0 \\
  1 \\
\end{array}%
\right), x<0 \\
c_{1}\exp(ik_{+}^{r})\left(%
\begin{array}{c}
  \cos(\theta/2)e^{-i\phi/2}  \\
  \sin(\theta/2)e^{i\phi/2}  \\
\end{array}%
\right)+
     d_{1}\exp(ik_{-}^{r})\left(%
\begin{array}{c}
  -\sin(\theta/2)e^{-i\phi/2}  \\
  \cos(\theta/2)e^{i\phi/2}  \\
\end{array}%
\right),x>L \\
     \end{array}%
 \right.
\end{equation}
In this equation,
$k_{\pm}^{l(r)}=\sqrt{2m(E-U_{l(r)}\mp{h_{l(r)}})/\hbar^{2}-k_{//}^{2}}$
is the spin-dependent wavevector along the $x$-direction in the
left (right) FM electrode, $E$ is the single-electron energy which
is conserved when electrons tunnel through the junction,
${\mathbf{k}}_{//}$ is the wavevector parallel to the interface
between the different regions and assumed to be conserved in the
quantum tunnelling process. Nevertheless, this is not a required
condition for the following derivation. $\Psi_{1}$ describes a
minority-spin electron coming from the left FM lead and being
scattered in the middle region; the scattering coefficients
$a_{1}$, $b_{1}$, $c_{1}$ and $d_{1}$ correspond to the normal
reflection, Andreev-type reflection,\cite{8} transmission without
branch crossing, and transmission with branch crossing,\cite{26}
respectively. For the other $3$ scattering processes $\Psi_{i}$ as
well as their coefficients $a_{i}$, $b_{i}$, $c_{i}$, and $d_{i}$
as shown in Fig.~2 have the same meaning . It is noted that we
have omitted the parallel plane wave component
$e^{i{\mathbf{k}}_{//}\vec{r}(y,z)}$ in the above equation. When
we use the continuity of the derivatives of wavefucntions at the
interfaces to determine these coefficients ( $a_i$, $b_i$, $c_i$,
and $d_i$), the parallel momentum ${\mathbf{k}}_{//}$ should be
explicitly taken into account.

With four elementary scattering wavefunctions like those in Eq.~2
above, the single-electron Green's function of the junction system
can be worked out by the McMillan formula,\cite{22} which has been
further developed by Kashiwaya and Tanaka.\cite{25} This formula
has been used to treat the Josephoson current in a superconductor
junction and relate directly the super current to the Andreev
reflection amplitudes. The Green's function $G^{r}(x,x')$ is
proportional to the direct product of the left-going wavefunctions
(processes $i=3$, $4$ in Fig.~2) and the right-going wavefunctions
(processes $i=1$, $2$), $G^{r}\sim \Psi_{L}(x)\hat{\Psi}_{R}(x')$
for $x\leq x'$ and $G^{r}\sim \Psi_{R}(x)*\hat{\Psi}_{L}(x')$ for
$x\geq x'$, where the hat `$\wedge$' denotes the conjugate process
to the elementary scattering one shown in Fig.~2, and these
conjugate scattering wavefunctions can be obtained by determining
the Hermitian conjugate of only the spinor part of the
wavefunction not including the spatial part of the wavefunction.
The reflection and transmission coefficients $\tilde{a}_i$,
$\tilde{b}_i$, $\tilde{c}_i$, and $\tilde{d}_i$  in four conjugate
processes have the relations $\tilde{a}_{i}(\phi)=a_{i}(-\phi)$,
$\tilde{b}_{i}(\phi)=b_{i}(-\phi)$,
$\tilde{c}_{i}(\phi)=c_{i}(-\phi)$, and
$\tilde{d}_{i}(\phi)=d_{i}(-\phi)$ ($i=1 \cdots 4$), where $\phi$
is the azimuthal angle of the magnetization of the right FM. With
these scattering wavefunctions of the elementary processes as well
as their conjugate processes, the Green's function is then
constructed in a linear combination as
\begin{equation}
G^{r}(x,x',E)=\left\{%
\begin{array}{l}
    \alpha_{1}\Psi_{3}(x)\hat{\Psi}_{1}(x')+\alpha_{2}\Psi_{3}(x)\hat{\Psi}_{2}(x')
   +\alpha_{3}\Psi_{4}(x)\hat{\Psi}_{1}(x')+\alpha_{4}\Psi_{4}(x)\hat{\Psi}_{2}(x'), x\leq x' \\
    \beta_{1}\Psi_{1}(x)\hat{\Psi}_{3}(x')+\beta_{2}\Psi_{2}(x)\hat{\Psi}_{3}(x')
   +\beta_{3}\Psi_{1}(x)\hat{\Psi}_{4}(x')+\beta_{4}\Psi_{2}(x)\hat{\Psi}_{4}(x'), x\geq x' \\
\end{array}%
\right.
\end{equation}
Here $G^{r}(x,x',E)$ is implicitly a function of the parallel
momentum $k_{//}$. The prefactors $\alpha_{i}$ and $\beta_{i}$
($i=1\cdots 4$) can be determined by the boundary conditions that
the Green's function fulfills,
\begin{equation}
\frac{\partial}{\partial
x}G^{r}(x,x',E)|_{x=x'+0}-\frac{\partial}{\partial
x}G^{r}(x,x',E)|_{x=x'-0}=\frac{2m_{e}}{\hbar^2}\left(%
\begin{array}{cc}
  1 & 0 \\
  0 & 1 \\
\end{array}%
\right),
\end{equation}
\begin{equation}
G^{r}(x,x',E)|_{x=x'+0}=G^{r}(x,x',E)|_{x=x'-0}.
\end{equation}
With these two equations, we can directly solve the prefactors
$\alpha_{i}$ and $\beta_{i}$ in Eq.~3, which are independent of
the spatial position $x$. After some direct algebra, they read
\begin{eqnarray}
\alpha_{1}=\frac{z_{+}c_{4}}{c_{3}c_{4}-d_{3}d_{4}},\ \
\alpha_{2}=\frac{-z_{-}d_{4}}{c_{3}c_{4}-d_{3}d_{4}}, \nonumber \\
 \alpha_{3}=\frac{-z_{+}d_{3}}{c_{3}c_{4}-d_{4}d_{4}},\ \
 \alpha_{4}=\frac{z_{-}c_{3}}{c_{3}c_{4}-d_{4}d_{4}}, \nonumber \\
\beta_{1}=\frac{z_{+}\tilde{c}_{4}}{\tilde{c}_{3}\tilde{c}_{4}-\tilde{d}_{3}\tilde{d}_{4}},\
\
\beta_{2}=\frac{-z_{-}\tilde{d}_{4}}{\tilde{c}_{3}\tilde{c}_{4}-\tilde{d}_{3}\tilde{d}_{4}}, \nonumber \\
\beta_{3}=\frac{-z_{+}\tilde{d}_{3}}{\tilde{c}_{3}\tilde{c}_{4}-\tilde{d}_{3}\tilde{d}_{4}},\
\
\beta_{4}=\frac{z_{-}\tilde{c}_{3}}{\tilde{c}_{3}\tilde{c}_{4}-\tilde{d}_{3}\tilde{d}_{4}},
\end{eqnarray}
where $z_{\pm}=\frac{m_{e}}{i\hbar^{2}k_{\pm}^{l}}$. In these
solutions, we have employed some detailed balance conditions to
facilitate our derivation, such as $a_{i}(\phi)=a_{i}(-\phi)$
($i=1\cdots 4$) and $k_{-}^{l}\tilde{b}_{1}=k_{+}^{l}b_{2}$.
Substituting these prefactors into Eq.~3, we obtained the Green's
function in the left FM electrode ($x$,$x'<0$) as
\begin{eqnarray}
G^{r}(x,x',E)=\left(%
\begin{array}{cc}
  G_{11} &   G_{12} \\
  G_{21} &   G_{22} \\
\end{array}%
\right), \nonumber \\
G_{11}=z_{+}\exp(ik_{+}^{l}|x-x'|)+a_{1}z_{+}\exp(-ik_{+}^{l}x-ik_{+}^{l}x'),\nonumber \\
G_{12}=z_{-}b_{2}\exp(-ik_{+}^{l}x-ik_{-}^{l}x'),\nonumber \\
G_{21}=z_{+}b_{1}\exp(-ik_{+}^{l}x'-ik_{-}^{l}x),\nonumber \\
G_{22}=z_{-}\exp(ik_{-}^{l}|x-x'|)+a_{2}z_{-}\exp(-ik_{-}^{l}x-ik_{-}^{l}x').\nonumber \\
\end{eqnarray}
The Green's functions in other regions such as the middle region
between the two FMs can also be constructed in a similar manner
according to Eq.~3 with solutions given in Eq.~6. Although the
Green's function is obtained here for the FM/FM junction, this
method can also be applied to nonmagnetic junctions, and we
believe the derivation procedure is universal in mesoscopic
junction systems. When the retarded Green's function of the
studied system is worked out, we can in principle calculate the
physical observables that we want. Since we focus on an
equilibrium tunnelling junction, the lesser Green's function is
easily obtained by the formula
\begin{equation}
G^{<}(x,x',E)=\left[G^{a}(x,x',E)-G^{r}(x,x',E)\right]f(E)
\end{equation}
where $G^{a}(x,x',E)$ is the advanced Green's function and $f(E)$
is the Fermi-Dirac distribution function. The lesser Green's
function is related to the spectral function of the electron and is
very useful for calculating some physical quantities as shown later.

\section{ESC in the FM/FM junction}
In this section, we first present a general formula of the spin
current in an FM and then utilize the Green's function obtained in
Sec.~II to work out the ESC in a FM/FM tunnelling junction. Then
as an example, we calculated ESC of a simple FM/FM junction with a
delta function insulator barrier between the two FM electrodes.

The local spin density $\vec{s}({\mathbf{r}},t)$ at position
$\mathbf{r}$ and time $t$ is defined as
\begin{equation}
\vec{s}({\mathbf{r}},t)=\Psi^{\dagger}({\mathbf{r}},t)\hat{\vec{s}}\Psi({\mathbf{r}},t),
\end{equation}
where $\Psi({\mathbf{r}},t)$ is a two-component wavefunction and
$\hat{\vec{s}}=\frac{\hbar}{2}\hat{\vec{\sigma}}$ with $\hat{\vec{\sigma}}$
denoting the Pauli spin matrices. Taking the time-derivative of
$\vec{s}({\mathbf{r}},t)$ and using the Schrodinger equation for a
single FM, the continuity equation of spin current density is
given by
\begin{equation}
\frac{\partial \vec{s}}{\partial t}+\nabla\cdot
{\mathbf{J}}_{s}+{\mathbf{S}}=0,
\end{equation}
where the respective spin current density ${\mathbf{J}}_s$ and the
source term $\mathbf{S}$ are
\begin{eqnarray}
{\mathbf{J}}_s=\frac{\hbar^{2}}{4m_{e}i}\left[\Psi^{\dagger}{\boldsymbol
{\sigma}}\nabla\Psi- (\nabla\Psi)^{\dagger}{\boldsymbol
{\sigma}}\Psi\right],\nonumber \\
{\mathbf{S}}=\Psi^{\dagger}({\boldsymbol\sigma}\times{\mathbf{h}})\Psi,
\end{eqnarray}
where $\mathbf{h}$ is the magnetization of the FM leads (for
example, ${\mathbf{h}}_{l}$, ${\mathbf{h}}_{r}$). ${\mathbf{J}}_s$
is the usual definition of spin current density and $\mathbf{S}$
is referred to as the spin source term or spin torque that leads
to spin rotation when the spin is not collinear with the
magnetization $\mathbf{h}$. Therefore this spin source term gives
rise to a spin current flow which is not included in
${\mathbf{J}}_s$. Using the lesser Green's function in Eq.~8,
$G^{<}(xt,x't')=i\langle\Psi^{\dagger}(x't')\Psi(xt)\rangle$ where
$\langle\cdots\rangle$ is the quantum statistical average, the
spin current density ${\mathbf{J}}_s$ and the source term
$\mathbf{S}$ above in a steady state can be rewritten as
\begin{equation}
{\mathbf{J}}_s=\frac{-\hbar^{2}}{4m_{e}}\lim_{x'\rightarrow
x}\left\{\frac{\partial}{\partial x'}-\frac{\partial}{\partial
x}\right\}\int\frac{dE}{2\pi}Tr\left[{\boldsymbol\sigma}
G^{<}(x',x)\right],
\end{equation}
\begin{equation}
{\mathbf{S}}=\lim_{x'\rightarrow
x}\int\frac{dE}{2\pi}Tr\left[({\mathbf{h}}\times{\boldsymbol\sigma})
G^{<}(x',x)\right],
\end{equation}
where the trace is taken over the spin space. Substituting the
Green's function given by Eq.~7 and Eq.~8 into the above equation,
we obtain the ESC expression in the FM/FM junction as
\begin{mathletters}
\begin{eqnarray}
J_{s}^{x}=\int\frac{dE}{4\pi}\sum_{k_{//}}
Re\left\{(k_{+}^{l}-k_{-}^{l})(\frac{b_{2}}{k_{-}^{l}}-\frac{b_{1}}{k_{+}^{l}})\exp[-i(k_{+}^{l}+k_{-}^{l})x]
\right\}f(E),
\end{eqnarray}
\begin{eqnarray}
J_{s}^{y}=-\int\frac{dE}{4\pi}\sum_{k_{//}}
Im\left\{(k_{+}^{l}-k_{-}^{l})(\frac{b_{2}}{k_{-}^{l}}+\frac{b_{1}}{k_{+}^{l}})\exp[-i(k_{+}^{l}+k_{-}^{l})x]
\right\}f(E)
\end{eqnarray}
\begin{eqnarray}
J_{s}^{z}=0.
\end{eqnarray}
\end{mathletters}
Here the summation over $k_{//}$ includes the contribution of all
transverse modes to the spin current, where $k_{//}$ is assumed to
be conserved in the quantum tunnelling process. According to the
above equation, the spin current density is determined by the
Andreev-type reflection coefficients $b_{1}$ and $b_{2}$; this
result is very similar to the formula of the Josephson current
which is directly related to the Andreev reflection
coefficients.\cite{24,25} The spin current density in Eq.~14 is
exponentially $x$-dependent so that the integral over energy and
transverse density of states will make it disappear in bulk FM not
far away from interface. This phenomena is same as the spin
transfer effect that the transmitted spin current is also
dependent on position and will vanish after summing all possible
transverse modes with different wavevectors. Stiles and
Zagwill\cite{6} estimated the characteristic length of this
spatial spin precession was about $1/k_f$ ($k_{f}$ Fermi
wavevector) because only the electrons near the Fermi energy
contribute to the electric current. For our case in Eq.~14, all
the band electrons contribute to ESC so that ${\mathbf{J}}_s$ is
expected to decay much more quickly with distance from interface.

Apart from the spin current density ${\mathbf{J}}_s$, which can be
calculated directly from its definition, additional spin current flow arises from the source
term $\mathbf{S}$ 
according to Eq.~10. According to the Gaussian theorem, another spin
current ${\mathbf{J}}_{soc}$ can be defined from this source term as
\begin{mathletters}
\begin{eqnarray}
J_{soc}^{(x)}=\int_{x}^{0}S^{x}dx
 =\int\frac{dE}{4\pi}\sum_{k_{//}}
Re\left\{(k_{+}^{l}-k_{-}^{l})(\frac{b_{2}}{k_{-}^{l}}-\frac{b_{1}}{k_{+}^{l}})(1-\exp[-i(k_{+}^{l}+k_{-}^{l})x]
)\right\}f(E)
\end{eqnarray}
\begin{eqnarray}
J_{soc}^{y}=\int_{x}^{0}S^{y}dx=-\int\frac{dE}{4\pi}\sum_{k_{//}}
Im\left\{(k_{+}^{l}-k_{-}^{l})(\frac{b_{2}}{k_{-}^{l}}+\frac{b_{1}}{k_{+}^{l}})(1-\exp[-i(k_{+}^{l}+k_{-}^{l})x]
)\right\}f(E).
\end{eqnarray}
\end{mathletters}
The $z$-component of ${\mathbf{J}}_{soc}$ is zero. The
$x$-dependent part of ${\mathbf{J}}_{soc}$ is just the
$-{\mathbf{J}}_{s}$ in Eq.~14, so that the sum of ${\mathbf{J}}_{soc}$ and ${\mathbf{J}}_{s}$ is independent of $x$ and keep constant from the
interface to the bulk FM, which does not disappear away from the
interface,
\begin{mathletters}
\begin{eqnarray}
J^{x}=\int\frac{dE}{4\pi}\sum_{k_{//}}
Re\left\{(k_{+}^{l}-k_{-}^{l})(\frac{b_{2}}{k_{-}^{l}}-\frac{b_{1}}{k_{+}^{l}})\right\}f(E),
\end{eqnarray}
\begin{eqnarray}
J^{y}=-\int\frac{dE}{4\pi}\sum_{k_{//}}
Im\left\{(k_{+}^{l}-k_{-}^{l})(\frac{b_{2}}{k_{-}^{l}}+\frac{b_{1}}{k_{+}^{l}})\right\}f(E).
\end{eqnarray}
\end{mathletters}
When a spin current ${\mathbf{J}}_{s}$ enters the bulk FM, it will
be absorbed by the lattice in FM because the magnetization $\mathbf{h}$
rotates the spin and the sum of ${\mathbf{J}}_{soc}$ and ${\mathbf{J}}_{s}$ therefore remains constant.

As an example, we calculated a simple case of the FM/FM tunnelling
junction, the potential energy in both FM is $U_{l}=U_{r}=0$, and
the insulator barrier is described by a delta barrier
$U_{0}\delta(x)$ where $U_{0}$ denotes the strength of the
barrier, The magnitudes of the magnetizations in two FM are equal
$h_{l}=h_{r}$. Using simple quantum mechanics method, we obtain
the Andreev-type reflection amplitudes as
\begin{eqnarray}
b_{1}=2(k_{-}-k_{+})k_{+}\sin\theta\exp(i\phi)/A,\nonumber \\
b_{2}=2(k_{-}-k_{+})k_{-}\sin\theta\exp(-i\phi)/A, \nonumber \\
A=k_{-}^{2}+k_{+}^{2}+6k_{+}k_{-}-2k_{0}^{2}+4ik_{0}(k_{+}+k_{-})-(k_{+}-k_{-})^2\cos\theta,
\end{eqnarray}
where $k_{\pm}=k_{\pm}^{l(r)}$ and $k_0=mU_{0}/\hbar^2$.
Substituting these quantities into Eq.~16, the total ESC is then
given by
\begin{equation}
{\mathbf{J}}=\int\frac{dE}{4\pi}\sum_{k_{//}}Im[1/A(E)](k_{+}-k_{-})f(E)
\hat{\mathbf{h}}_{l}\times\hat{\mathbf{h}}_{r}
\end{equation}
where $\hat{\mathbf{h}}_{l}(0,0,1)$ and
$\hat{\mathbf{h}}_{r}(\sin\theta\cos\phi,\sin\theta\sin\phi,\cos\theta)$
are just the unit vectors of the magnetizations in the left and
right FM. The ESC depends on the directions of the two
magnetizations not only through the cross product term
$\hat{\mathbf{h}}_{l}\times\hat{\mathbf{h}}_{r}$ but also on the
quantity $A$ in Eq.~17. In the finite barrier case, we can neglect
the $\cos\theta$ term in $A$ so that Eq.~18 can reproduce the
result obtained with the linear response theory, i.e., the
exchange coupling between two magnetization leading to ESC. At the
opposite limit, zero barrier between two FMs $U_{0}=0$, Eq.~18
indicates there is no ESC flowing in the junction. This is the
pure quantum mechanics effect, the reflected spin direction by a
barrier rotates with respect to the incident spin direction and an
imaginary part can appear in the quantity $A$ of Eq.~17, which was
regarded as\cite{6} one of reasons that spin current is not
conserved at the interface when a polarized charge current flows
through a NM/FM junction. For ESC discussed here, the reflections
by a barrier leading to an additional phase to the incident
wavefunction which is necessary for the formation of ESC in the
noncollinear FM/FM junction. It is quite different from the
Josephson effect between two superconductors, in which a cooper
pair carries supercurrent through the junction even with zero
barrier.

\section{summary}
We have studied ESC flowing in the noncollinear FM/FM tunnelling
junction by extending McMillan Green's function method. The
single-electron Green's function of the junction can be
constructed by the scattering coefficients, reflection and
transmission amplitudes which can be directly calculated by simple
quantum mechanics method. In the derived formula of spin
current density, we found that the exact result of ESC does not
have a big difference from that obtained by the linear response
theory, in other words, the high-ordered tunnelling process
contribute little to ESC. It was also found that
when there is no barrier between two FMs, the ESC will disappear for the
reflected spin state from a finite barrier has an additional phase
which is crucial for the formation of ESC.

ACKNOWLEDGEMENT  This work is supported by City University of
Hong Kong Strategic Research Grant (Project No. 7001619).

\newpage

\begin{figure}
\caption{Schematic of a FM/FM tunnelling junction with two
noncollinear magnetizatons in the left and right FM electrodes. The middle region
M at $0<x<L$ can be either an insulator barrier or a normal metal.
The $z$-direction is taken as the spin quantization axis parallel
to the left magnetization ${\mathbf{h}}_{l}$.}
\end{figure}

\begin{figure}
\caption{A schematic picture of four elementary scattering events.
Every injecting electron will be scattered into four outgoing wave
functions, two reflections and two transmissions, e.g.,
$\Psi_{1}(x)$ denotes a minority spin electron $k_{+}^{l}$ in the left
FM injecting into the middle region with two reflection ($a_{1}$
and $b_{1}$) and two transmission ($c_{1}$ and $d_{1}$)
wavefunctions. }
\end{figure}

\end{document}